# GNSS-R: Operational Applications

G. Ruffini, O. Germain, F. Soulat, M. Taani and M. Caparrini

*Starlab, Edifici de l'Observatori Fabra, 08035 Barcelona, Spain,* <http://starlab.es>

**ABSTRACT**

This paper provides an overview of operational applications of GNSS-R, and describes Oceanpal®, an inexpensive, all-weather, passive instrument for remote sensing of the ocean and other water surfaces. This instrument is based on the use of reflected signals emitted from GNSS, and it holds great potential for future applications thanks to the growing, long term GNSS infrastructure. The instrument exploits the fact that, at any given moment, several GNSS emitters are simultaneously in view, providing separated multiple scattering points with different geometries. Reflected signals are affected by surface "roughness" and motion (i.e., sea state, orbital motion, and currents), mean surface height and dielectric properties (i.e., salinity and pollution). Oceanpal® is envisioned as an accurate, "dry" tide gauge and surface roughness monitoring system, and as an important element of a future distributed ocean remote sensing network concept. We also report some results from the Starlab Coastpal campaign, focusing on ground GNSS-R applications.

**INTRODUCTION**

Several GNSS constellations and augmentation systems are presently operational, such as the Global Positioning System (GPS), owned by the United States, and, to some extent, the Russian GLObal NAvigation Satellite System (GLONASS). In the next few years, the European Satellite Navigation System (Galileo) will be deployed. By the time Galileo becomes operational in 2008, more than 50 GNSS satellites will be emitting very precise L-band spread spectrum signals, and will remain in operation for at least a few decades. Although originally meant for localization, these signals will no doubt be used within GCOS/GOOS. The immediate objective of Starlab's Oceanpal® project is the development of technologies for operational in-situ or low-altitude water surface monitoring using GNSS Reflections, a passive, all weather radar technology of great potential.

Oceanpal® is an offspring of technology developed within several ESA/ESTEC projects targeted on the exploitation of GNSS Reflections from space[1], following the proposal of M. Martín-Neira (1993). We also note that GNSS-R is but an example of passive, bistatic radar (see, e.g., Cantafio 1993), a subject with a long history. In fact, bistatic radar was a subject of research in the early days—see Conant (2002) for a fascinating account of radar history.

Although our focus here is on low altitude applications, it is worthwhile explaining in more detail the rationale for spaceborne deployment: an important aspect of the GNSS-R concept is the synergy between space and ground monitoring using the same technology and the same signal infrastructure, which will ensure homogeneity in the measurements. An overview of the parameters measured by GNSS-R is provided in Table 1.

In Figure 1 we can see a schematic rendition of a spaceborne GNSS-R mission, as well as an illustration showing the multiple (GPS) reflection points available to a ground receiver during a 24-hour period. Note the multi-static character of the technique: a single passive instrument can provide a rather large swath, thanks to the availability of multiple emitters. From the ground and air, it can also provide simultaneous measurements in different geometric configurations over the same area—an important added value for geophysical inversion.

---

[1] Such as the ESA projects OPPSCAT, OPPSCAT 2 (focusing on Speculometry/Scatterometry), Paris-Alpha, Paris-Beta, Paris-Gamma (Altimetry) and GIOS-1 (focusing on Ionospheric monitoring). See the Acknowledgements for more details.

|        | *SWH* | *DMSS* | *H* |
|--------|:-----:|:------:|:---:|
| GROUND | ●     |        | ●   |
| AIR    |       | ●      | ●   |
| SPACE  |       | ●      | ●   |

**Table 1** Summary of the main measurements of GNSS-R for oceanography. Other possibilities include Surface Currents, Surface Pressure (from space) and Dielectric constant.

## 2. GNSS-R IN SPACE: THE PETREL EARTH EXPLORER

In the future, the artificial separation between geophysical "layers" (ocean, troposphere, stratosphere, etc.) will disappear, and future Earth global models will need to reflect the fundamental role of atmosphere-ocean coupling. The sea surface provides the ocean-atmosphere link, regulating momentum, energy and gas exchange, and several fundamental ocean circulation features are directly related to wind-wave induced turbulent transports in the oceanic mixed layer. In particular, eddies and gyres are fundamental agents for mixing, heat transport and feedback to general circulation, as well as transport of nutrients, chemicals and biota for biochemical processes.

Moreover, at the atmosphere-ocean boundary, many temporal and spatial scales play an important role: from the molecular to the synoptic level, from seconds to eons. For this reason, observing this interface appropriately is an important challenge for global observation systems, which will require high resolution, wide swaths, frequent revisits and long-term stability (Le Traon et al, 2002). All of these are actively addressed by the GNSS-R concept.

The ocean-atmosphere interface is characterized (to the lowest statistical order) by the geophysical variables of local mean sea level ($h$), significant wave height ($swh$) and directional mean square slope ($dmss$). Mesoscale measurements of sea surface $dmss$ are an important missing element from the global climate and ocean observation systems, and would greatly help to understand and quantify the atmosphere-ocean flux of energy, momentum and gas. In addition, since ocean forcing is a non-linear and strongly intermittent phenomenon (both in space and time), frequent space-time co-located mesoscale measurements of $h$ and $dmss$ are highly desirable. A similar statement, asserting the importance of simultaneous altimetry and scatterometry measurements, was already stated in 1981 (WOCE CCCO, see Thompson et al., p. 35 in Siedler et al., 1981).

The scientific objectives of a spaceborne GNSS-R mission such as PETREL, recently submitted to the Earth Explorer ESA program (Ruffini and Chapron, 2002) should thus address the medium and long-term components for physical climate observation (Theme 2 of the ESA Earth Explorer Program) with a focus on providing a key element for the study of atmosphere-ocean coupling.

The elementary geophysical products provided by such a mission highlight mesoscale collocated altimetric and sea surface directional mean square slope measurements.

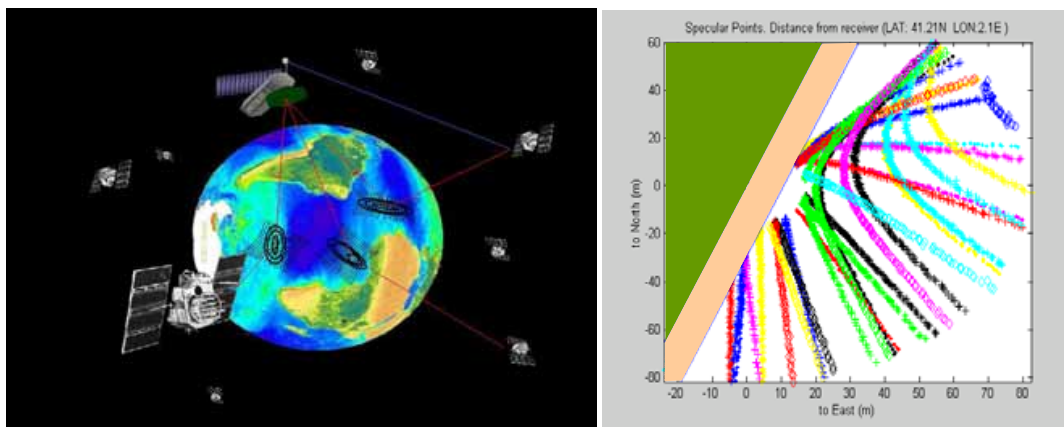

**Figure 1:** <u>Left:</u> Artist concept of the bistatic GNSS spaceborne concept. GNSS signals reflect off the Earth surface and are gathered by a spaceborne receiver. All direct signal links are not shown for simplicity. <u>Right:</u> GPS-R specular points after 24 hours as seen from a receiver at 50 m altitude in the Barcelona coast. A conservative cut-off of 20 degrees in the local elevation of the reflected signals has been used for display purposes.

These measurements are also of great interest for the observation of surface winds, mean sea surface, sea-ice, salinity, ionospheric electron content and tropospheric delay (e.g., for measurement of surface pressure over the oceans). The measurement of currents is in principle also feasible.

Results from recent ESA studies and experiments and also from our colleagues in the United States indicate that GNSS-R data can provide sufficient information to resolve mesoscale features in the ocean, as well as co-located directional mean square slope measurements (see, e.g., other papers from the *2003 Workshop on Oceanography with GNSS Reflections* in the references).

As recent ESA studies indicate, a single GNSS-R Low Earth Orbiter can provide samplings of less than 100 km resolution and less than 10 days revisit time with an equivalent altimetric precision better than 5 cm, sufficient for competitive Mesoscale Altimetry applications. Such a mission, capable of picking as many as 12 signals from GPS, Galileo and Inmarsat satellites would have significant impact on the mapping of the mesoscale variability.

According to the present understanding of the GNSS-R error budget (based on theoretical studies and related experimental campaigns carried in Europe and the US), impact studies with simulated data carried out within the scope of the Paris Beta and Paris Gamma ESA studies show that a GNSS-R mission should allow mapping of the mesoscale variability in high eddy variability regions better than Jason-1+ENVISAT together. These studies also indicate that the combination of GNSS with Jason-1 and ENVISAT data can improve the sea level mapping derived from the combination of Jason-1 and ENVISAT by a factor of about 2. In well sampled regions, the improvement could reach up to a factor of 4 (see Le Traon et al., 2003).

The precision and sampling provided by such measurements may also make GNSS-R an effective tool for Tsunami detection and the measurement of surface pressure over the oceans (e.g., in the Southern Hemisphere or inside hurricanes). We recall that the troposphere induces a delay in GNSS signals which can be parameterized, to first order, by a measurement of surface pressure.

**RECENT COASTAL EXPERIMENTAL CAMPAIGNS**

Many experiments have taken place to date, carried out by different institutions in Europe and the US: from space, stratospheric balloons, aircraft and the ground. The reader is invited to read through the references for abundant experimental work. Here we report briefly on Starlab's 2003 Coastpal campaign.

Recent coastal GNSS-R experimental campaigns led by Starlab have collected data from low altitude stationary platforms in a wide range of sea state conditions, using both experimental GPS-R equipment lent by ESA/ESTEC and an Oceanpal® prototype. Some of these experiments (the Coastpal series) have been carried out in the Barcelona harbor breakers, with the logistic support of the Barcelona Port Authority.

As shown in Figure 2, two antennas are usually employed to collect GPS signals: one antenna (the "direct" or "up-looking" antenna) is zenith looking and Right Hand Circularly Polarized to collect the direct GPS signal, while the other one (the "reflected" or "down-looking" antenna) is nadir/side looking and Left Hand Circularly Polarized to recover the reflected signal. The output from each the antenna is sent to a GPS front end. The IF data generated by the receivers is then recorded at a sufficiently high sampling frequency, after (typically) being digitized at one bit. The experimental data has been fed to Starlab's GPS-Reflections processor (STARLIGHT[2]), which retrieves the reflected electromagnetic field and estimates sea level and sea state.

The STARLIGHT processor, through the conventional correlation method, evaluates the reflected field magnitude and phase. The retrieved field contains very useful information on the characteristics of the reflecting surface. Comparison between this field and the direct one is then performed to infer the desired quantities, such as sea roughness and sea level. Recent altimetric results using the phase in mild sea conditions are at the centimeter level (Caparrini et al., 2003), and there appears to be very good correlation between sea state and field dynamics.

Figure 2 shows some details of the experimental hardware set-up. The particular experiment shown in Figure 2 took place at dawn. Along with the GPS signals another source of opportunity was exploited: the Mediterranean rising sun. The use of multi-frequency bistatic specular scattering instruments is very important to validate models, and may provide clues on how to separate ocean surface spectral parameters (such as surface wind and wave age).

To understand the geophysical content of the data it is useful to perform a "gedanken[3]". As seen from a static platform, the electric field scattering from a frozen ocean could be represented as a static complex phasor (representing the phase and amplitude of the electric field). The reader can then readily imagine that the motion of the ocean surface translates into motion of the phasor in the complex plane.

In Figure 4 we can see such a phasor representation of the reflected electric field (at GPS frequencies) simulated using Fresnel scattering from an virtual ocean generated using the Elfouhaily et al. (1997) ocean spectrum, as well as the real thing obtained using experimental data from GPS L1 signals (from the Coastpal campaign in the Barcelona harbor). Analysis of the dynamics of the reflected phasor provides the key to estimating sea surface parameters from such static platforms. In Figure 3, for instance, we show some results from a study carried out with the help of the scattering simulator for sea state retrieval from phase statistics (Soulat, 2003). It also shows preliminary results obtained through a Fourier analysis of the complex reflected field gathered at the Barcelona Port for different sea conditions. As observed and as expected from simulations and analytic work, the energy and width of the spectrum increase quite clearly with surface wind speed, which is a very promising indication for the development of our inversion algorithms.

---

[2] STARLab Interferometric GNSS Toolkit.

[3] "Thought experiment", in German.

Aircraft or spacecraft observations must be analyzed differently, basically exploiting the size and shape of the "glistening" zone (as in Spooner, 1822, or in the classic work by Cox and Munk using optical data). The fundamental tool to study this is provided by the Delay Doppler mapping SAR-like capability of GNSS-R (see, e.g., Ruffini 1999 and 2000a, Germain 2003 and Soulat 2003).

## GNSS-R AS TIDE GAUGE AND SEA STATE SENSOR: THE OCEANPAL® CONCEPT

Starlab is now developing an operational instrument based on GNSS-R, Oceanpal®.

As we have seen, initial results indicate that this sensor will provide very useful altimetry and sea state information from, at least, low altitude applications (e.g., coasts or aircraft).

The company is perfecting robust algorithms for operational code and phase tracking of the reflected field and extraction of geophysical parameters. As discussed, reflected signals carry significant information on sea state and topography, and both experimental work and simulations have demonstrated the potential of this concept for coastal and airborne altimetry and sea state monitoring.

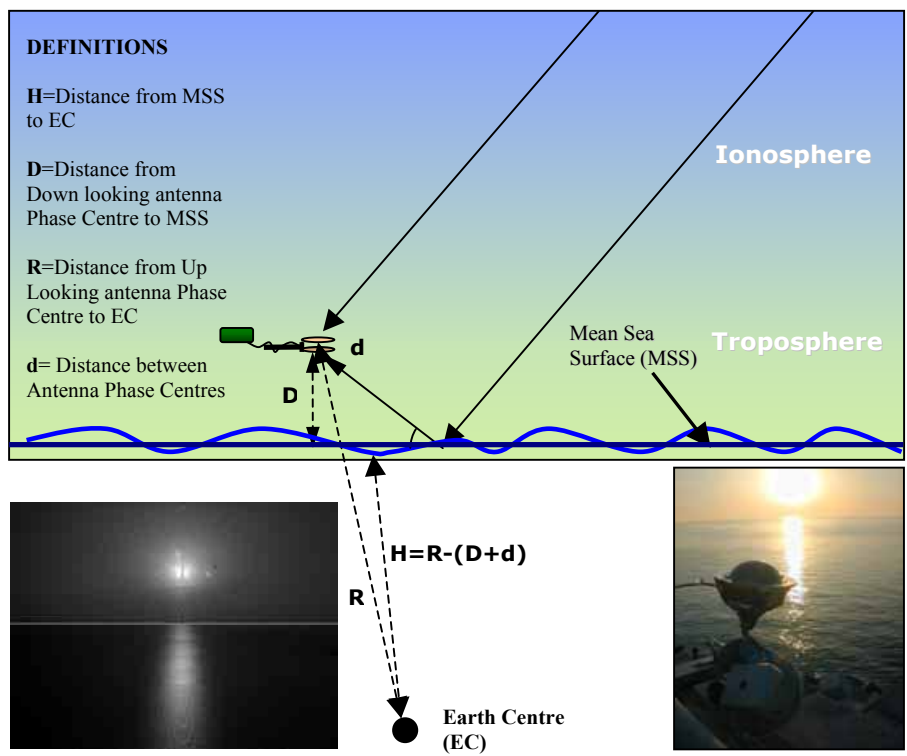

**Figure 2:** Simplified schematic representation of the GNSS-R concept. The direct and reflected signals originating from a GNSS source are combined at the receiver to estimate the distance to the surface and to the Earth Centre (atmospheric errors cancel out at low altitudes). <u>On the bottom right</u>, a detail of the Coastpal Experiment in the Barcelona Port using equipment provided by ESA-ESTEC and with the support of the Barcelona Port Authority. The up and down looking antennas can be seen. <u>On the bottom left</u>, example of analysis of optical glitter to support GNSS-R analysis. Slope statistics can be calculated using optical measurements.

As seen from the instrument, several GNSS emitters are simultaneously in view at any given time, providing information from separated scattering points with different geometries and thus strengthening the extraction of oceanographic variables (geophysical inversion). Reflected signals are affected by surface "roughness", motion (sea state, orbital motion, currents), surface dielectric properties (i.e., salinity and pollution), and mean surface height. The instrument exploits the "noisy" reflected electric field to infer ocean, river, or lake surface properties, using robust techniques.

Although bistatic radar can work exploiting various sources of opportunity, GNSS are in many ways unique: GNSS-R altimetric products are very stable, long-term and could provide, automatically, absolutely calibrated mean sea level in the GNSS reference system. Thanks to its GNSS "pedigree", Oceanpal® is an inexpensive, all-weather, passive concept for remote sensing of the ocean and other water surfaces, for accurate provision of sea state and altimetry. The instrument is design so it can be deployed on multiple platforms: static (coasts, harbors, off-shore), and slow-moving (e.g., boats, floating platforms, buoys, stratospheric platforms, aircraft, etc.). Spaceborne application of GNSS-R requires further technology development, and is the subject of several ongoing ESA projects.

We envisage that this system will act as an accurate, distributed, "dry" tide gauge network while conducting surface scattering monitoring, providing a stable and precise service based on the growing long term GNSS infrastructure. As such, Oceanpal® is part of another Starlab concept in which small, multiple inexpensive sensors will exchange information to "synthesize" an extended remote sensing system and provide relevant oceanographic information to a whole array of end-users (GOOS, Public Authorities, harbors, shipping, fishing industry, off-shore mining, and in general to those conducting their activities in or near the sea).

**SUMMARY AND OUTLOOK**

GNSS-R is a budding new technology with a bright outlook. We foresee powerful applications for altimetry and scatterometry from ground, air and space using GNSS based bistatic radar technology: geophysical applications will clearly benefit from the precision, accuracy, abundance, stability and long-term availability of GNSS signals.

In this paper we have highlighted an inexpensive, passive, dry operational sensor concept for use on coastal platforms and aircraft, now under development at Starlab. This sensor will provide precise sea level information and sea state, and we believe it will occupy an important niche in operational oceanography and marine operations. Other marine applications of this technology (salinity, pollution, currents) are also being studied.

However, we emphasize that ESA and other agencies are currently working on the development of GNSS-R space sensors: recent studies indicate that GNSS-R data will have a significant altimetric and speculometric impact from space in conjunction with standard approaches. Mesoscale altimetry is an important target of recent studies, since one of the strongest assets of GNSS-R is the intense availability of reflected signals, which can provide very dense and accurate samplings. Speculometry can provide measurements of directional sea surface roughness, which can then be correlated with surface winds and sea state for operational applications as well as used directly for scientific studies of ocean-atmosphere coupling.

Given the growing GNSS availability and long-term outlook for GNSS service signals, the combination of GNSS-R data from air, ground and space can provide a long lasting oceanographic monitoring infrastructure for decades to come.

**Acknowledgements**

This work was partly supported by a Spanish Ministry of Science and Technology PROFIT project. We are also thankful for the support received in the context of several GNSS-R Starlab-ESA/ESTEC contracts: OPPSCAT (13461/99/NL/GD), the ongoing OPPSCAT 2 (3-10120/01/NL/SF), both dedicated to GNSS-R scatterometry (Speculometry), as well as ESA/ESTEC Contract 15083/01/NL/MM (PARIS BETA), ESA/ESTEC Contract No. 14285/85/nl/pb, Starlab CCN3-WP3 (PARIS ALPHA) and the ongoing ESA PARIS GAMMA project (all dedicated to the study of GNSS-R spaceborne altimetric applications).

Special thanks to ESA/ESTEC for allowing us to use their GPS-R experimental equipment, to the Barcelona Port Authority (J. Vilá) and Polytechnic University of Catalunya/TSC (A. Camps) for experimental logistic support during the Coastpal campaign, and to our partners in these ESA projects.

*All Starlab authors have contributed significantly; the Starlab author list has been ordered randomly.*

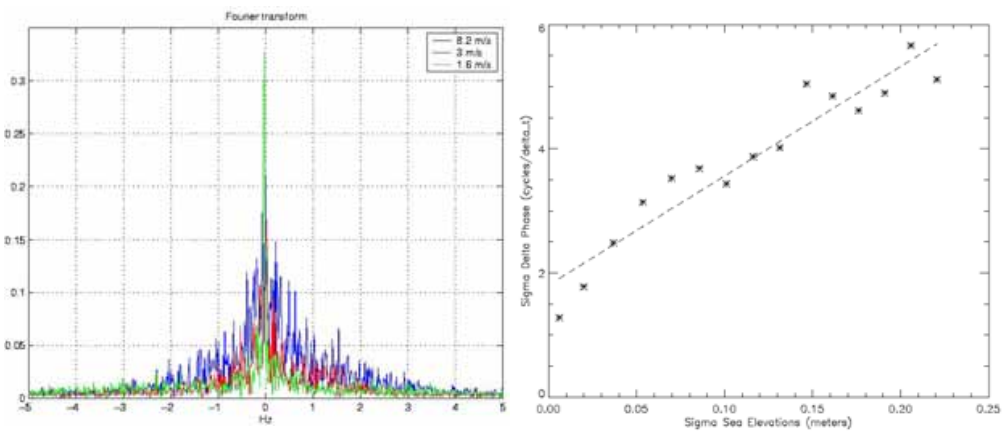

**Figure 3**. <u>Left:</u> Coastpal campaign data: Fourier analysis of the complex reflected field for different wind speeds: 1.6 m/s (green), 3 m/s (red) and 8.2 m/s (blue). <u>Right:</u> Simulations of phase dynamics statistics versus sea height RMS using Starlab's GNSS-R simulator (GRADAS).

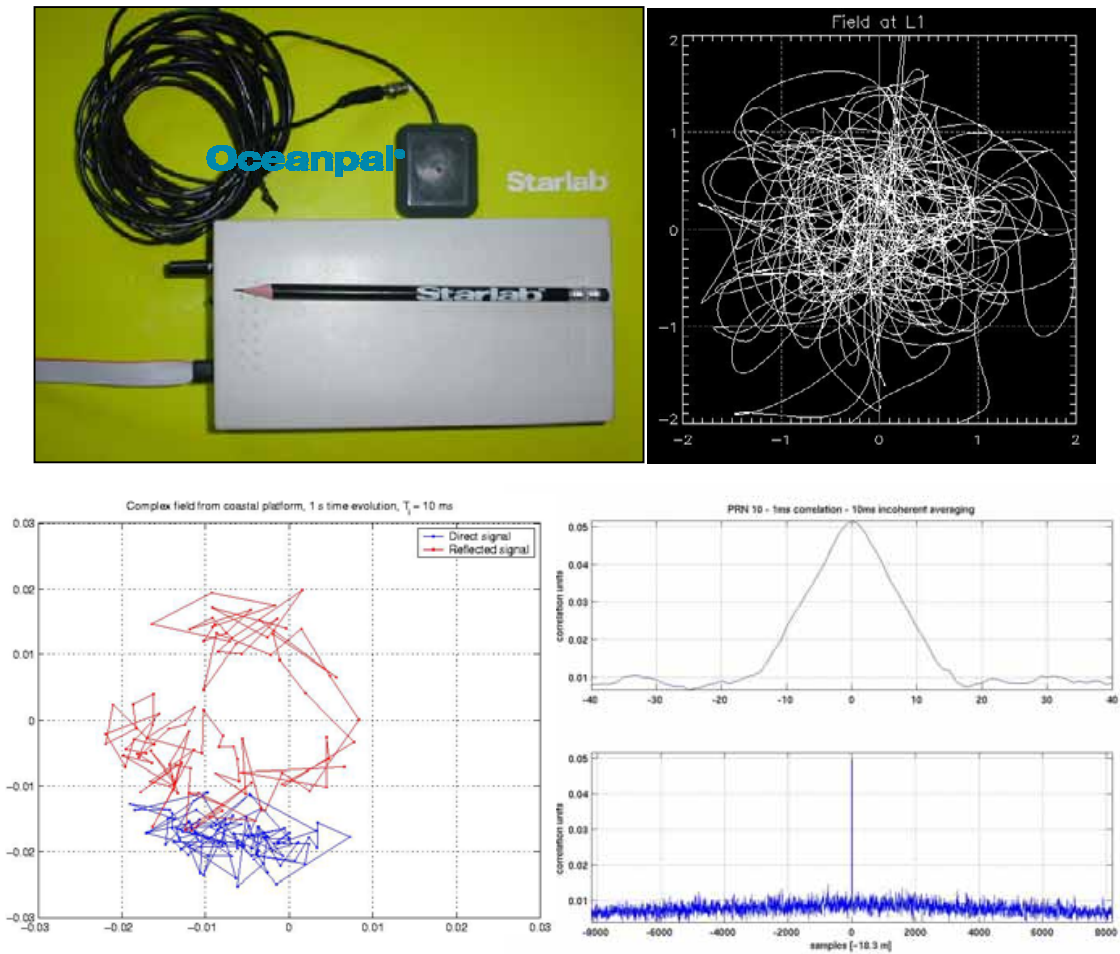

**Figure 4**. Top, left: Starlab's Oceanpal® prototype is a GNSS-R sensor ideally suited for coastal, river or lake applications. Top right: The dynamic L-band reflected electric field in a complex phasor representation of the amplitude and phase modulation of the carrier as produced by a virtual moving ocean after a few seconds of time evolution, from a simulation using Starlab's GRADAS software package (phasor amplitude units are arbitrary). Bottom: On the left, the dynamic phasor of the direct and reflected GPS L1 field after one second of time evolution, using data from the Coastpal experiment processed using Starlab's STARLIGHT software (units are SNRV, integration time is 10 ms). On the right, a typical Oceanpal® correlation waveform.

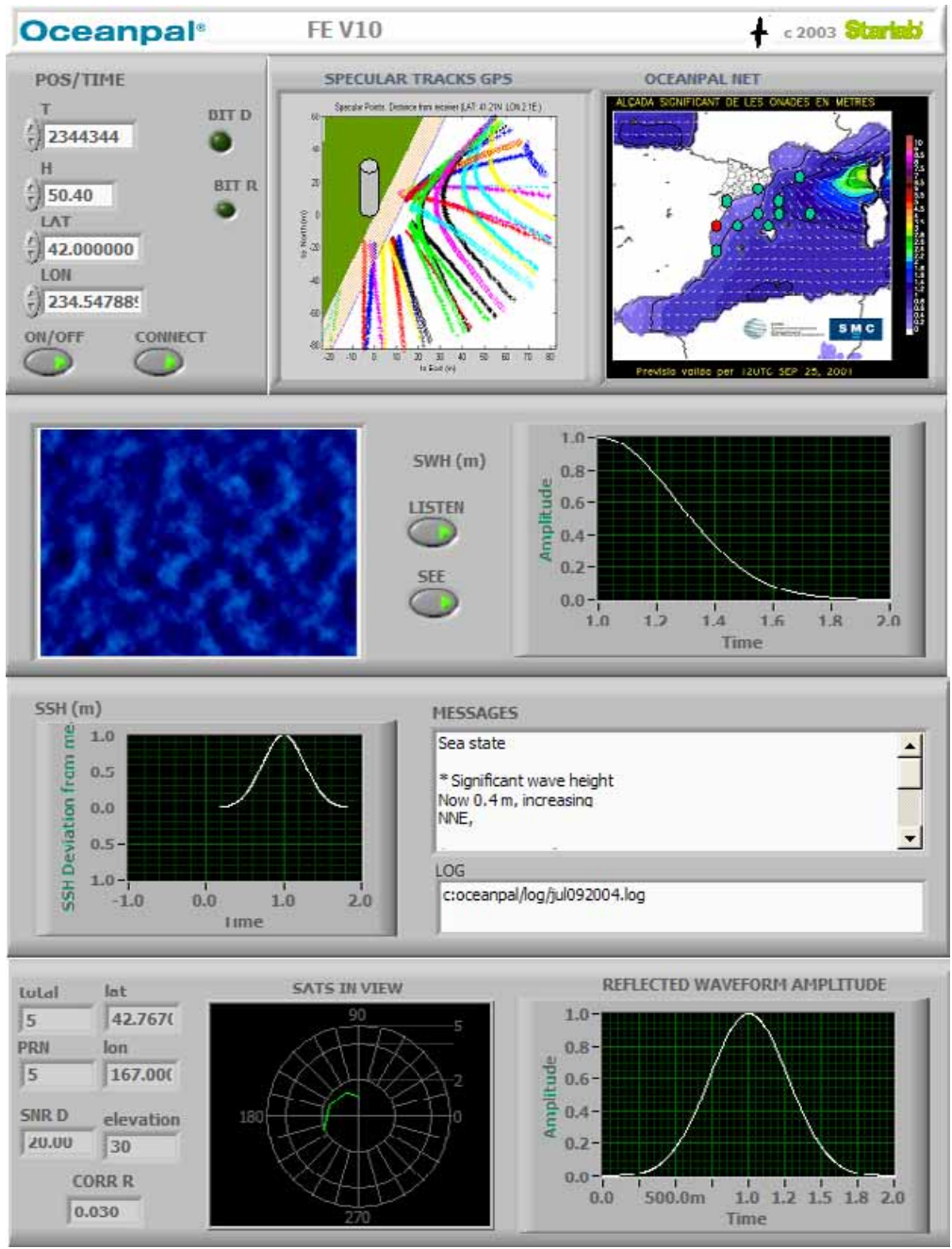

**Figure 5**. Oceanpal® interface concept. On the top panel, general information on the location of the sensor is provided, as well as on the available network of sensors and resulting overall sea state or Sea Surface Height map. On the second panel, the sea state (SWH index) is shown, as well as visual and acoustic cues on sea state. In the third panel, the Sea Surface Height is shown, as well as information in the form of text. Finally, information on the available satellites and signal "health" are provided.